\documentclass[aps,prl,twocolumn,showpacs,floatfix,superscriptaddress]{revtex4-1}

\usepackage[dvips]{graphics}
\usepackage{epsfig}
\usepackage{times}
\usepackage{amsmath,amssymb}
\usepackage{amsthm}
\usepackage{subfigure,overpic}
\usepackage{mathrsfs}
\usepackage{multirow}

\DeclareMathOperator{\Tr}{Tr}

\def \beq {\begin{equation}}
\def \edq {\end{equation}}
\def \bes {\begin{subequations}}
\def \eds {\end{subequations}}
\def \beqn {\begin{equation*}}
\def \edqn {\end{equation*}}

\begin{document}
\title{Scattering theory of nonlinear thermoelectric transport}
\author{David S\'anchez}
\author{Rosa L\'opez}
\affiliation{Institut de F\'{i}sica Interdisciplin\`{a}ria i de Sistemes Complexos
IFISC (UIB-CSIC), E-07122 Palma de Mallorca, Spain}
\affiliation{Departament de F\'{i}sica, Universitat de les Illes Balears,
E-07122 Palma de Mallorca, Spain}


\begin{abstract}
We investigate nonlinear transport properties of quantum conductors
in response to both electrical and thermal driving forces.
Within scattering approach, we determine the nonequilibrium
screening potential of a generic mesoscopic system
and find that its response is dictated by particle and entropic injectivities
which describe the charge and entropy transfer during transport.
We illustrate our model analyzing the voltage and thermal rectification
of a resonant tunneling barrier. Importantly, we discuss
interaction induced contributions to the thermopower in the presence
of large temperature differences.
\end{abstract}

\maketitle

\paragraph{Introduction.} Recent advances in nanoscale thermoelectric materials
suggest novel functionalities and highly improved performances \cite{row06}.
A key ingredient of thermoelectric devices is the Seebeck effect, which depends on the
simultaneous existence of thermal and electric driving forces. As a result,
energy conversion from waste heat is possible under the conditions of zero net
current. The Seebeck coefficient $S$ measures the amount of thermovoltage
generated across a conducting sample when a thermal gradient is externally applied.
Interestingly, the thermoelectric figure-of-merit is proportional to $S^2$. Therefore,
it is highly desirable to put forward new routes to increase $S$. Electron-electron
interactions may dramatically enhance $S$ in strongly correlated systems
as in magnetically diluted metallic hosts \cite{bic85} and artificial Kondo impurities \cite{cos10}.

On the other hand, large temperature drops give rise, quite generally, to
thermal rectification effects \cite{ter02}. The possibility to apply sharp thermal gradients
seems to be more feasible in nanostructured materias, as recently demonstrated
in superlattices with periods spanning a few nanometers \cite{ven01}.
Strikingly enough, a self-consistent theory of nonlinear thermoelectric
transport valid for quantum conductors is still lacking. This is the gap
we want to fill in this work.

Linear thermoelectric effects within the scattering approach were discussed
in Ref. \onlinecite{but90}. At the same time, pioneering experiments analyzed the main properties
of the thermopower at linear response in quantum point contacts \cite{mol92} and quantum dots \cite{sta93}.
Subsequent advances have unveiled fluctuating thermopower in chaotic dots \cite{god99},
large $S$ in Andreev interferometers \cite{eom98} and
thermoelectric anisotropies in multiterminal ballistic microjunctions \cite{mat12}.
The Seebeck coefficient can also help determine the conduction character of a molecular junction \cite{red07}.
Only recently has been possible a clear observation of thermal rectification
effects in mesoscopic systems \cite{sch08}. Thus, it is natural to ask
how phase-coherent current and thermopower are affected in the nonlinear
regime of transport.

In the isothermal case, all terminals are held at the same background temperature $T$.
Refs. \onlinecite{but93,chr96} then provide a convenient theoretical framework to include nonequilibrium
effects beyond linear response. The theory is based on an expansion around the equilibrium
point but, importantly, the nonlinear transport coefficients are complicated functions
of the screening response of the conductor out of equilibrium.
This purely interaction driven response is
described in terms of {\em characteristic potentials} which measure how the internal
potential counterbalance the ensuing charge pile-up due to a voltage shift.
Hence, the characteristic potentials depend on the {\em particle
injectivity} of those carriers originated in the shifted terminal.
The role of these particle injectivities is crucial because they
determine departures from the Onsager-Casimir symmetry relations \cite{san04,spi04}
ubiquitously found in nonlinear transport experiments \cite{mar06,let06,zum06,ang07,har08}.
Here, we show that when the system is
perturbed with a {\em temperature} shift its response is dictated
by {\em entropic injectivities} which quantify the entropy transported
in the charge imbalance process. Below, we discuss the role
of both particle and entropic injectivities in two
conceptually simple but generic problems---the formation of rectifying
terms in thermally driven electric currents and
the differential Seebeck coefficient beyond linear response.
\paragraph{Theoretical model.} 
We consider a mesoscopic conductor coupled to multiple terminals $\alpha,\beta\ldots$
characterized with bias voltages $eV_\alpha=\mu_\alpha-E_F$
($\mu_\alpha$ is the electrochemical potential and $E_F$ the Fermi energy)
and temperature shifts $\theta_\alpha=T_\alpha-T$
($T_\alpha$ is the reservoir temperature). The electronic transport is completely determined by
the scattering matrix $s_{\alpha\beta}=s_{\alpha\beta}[E,eU(\vec{r})]$ which, in general,
is a function of the carrier energy $E$ and the potential landscape inside the conductor $U(\vec{r})$ \cite{but93,chr96}.
In turn, $U(\vec{r})$ is a function of position $\vec{r}$ and the set of voltage
and temperature shifts. Defining $A_{\alpha\beta}=\Tr [\delta_{\alpha\beta}-
s_{\alpha\beta}^\dagger s_{\alpha\beta}]$, the electrical 
current is expressed as $I_\alpha=\frac{2e}{h}\sum_\beta\int dE A_{\alpha\beta}(E) f_{\beta}(E)$ where $f_\beta(E)=1/(1+\exp{[(E-E_F-eV_\beta)/k_B T_\beta]})$ 
is the Fermi distribution function in reservoir $\beta$.
In the weakly nonlinear regime of transport, the dominant terms
appear up to second order in an expansion of the electric current
in powers of the driving fields $V_\alpha$ and $\theta_\alpha$:
\begin{align}\label{I_exp}
I_\alpha &=\sum_{\beta}G_{\alpha\beta}V_\beta
+\sum_{\beta}L_{\alpha\beta}\theta_\beta
+\sum_{\beta\gamma}G_{\alpha\beta\gamma}V_\beta V_\gamma 
\nonumber \\
&+\sum_{\beta\gamma}L_{\alpha\beta\gamma}\theta_\beta \theta_\gamma
+2\sum_{\beta\gamma}M_{\alpha\beta\gamma}V_\beta \theta_\gamma\,.
\end{align}

The electrical and thermoelectric linear conductances are \cite{but90}
$G_{\alpha\beta}=-(2e^2/h)\int dE\,  A_{\alpha\beta} \,\partial_E f\simeq (2e^2/h) \,A_{\alpha\beta}(E_F)$
and $L_{\alpha\beta}=-(2e/hT)\int dE\, (E-E_F) A_{\alpha\beta}\, \partial_E f\simeq  (2e\pi^2 k_B^2 T/3 h)\, \partial_E A_{\alpha\beta}|_{E=E_F}$, respectively, where the approximate 
expressions correspond to a Sommerfeld expansion to leading order in $k_B T/E_F$.
Here, $f$ is the Fermi distribution function when all $V_\alpha$ and $\theta_\alpha$ are set to zero.
We emphasize that the linear conductances are evaluated at equilibrium and, as a consequence, $G_{\alpha\beta}$
and $L_{\alpha\beta}$ are independent of the screening potential $U$. The situation is completely different
for the nonlinear coefficients. We find,
\begin{subequations}\label{nonlinearcoefficients}
\begin{align}
G_{\alpha\beta\gamma}&=\!\frac{-e^2}{h}\!\! \int \! dE 
\left( \!\!\frac{\partial A_{\alpha\beta}}{\partial V_\gamma } \!+\!\frac{\partial A_{\alpha\gamma}}{\partial V_\beta } \!+\!e\delta_{\beta\gamma} \partial_E A_{\alpha\beta}\!\! \right)\! \partial_E f,\\
L_{\alpha\beta\gamma}&= \!\frac{e}{h} \!\int \!\! dE  \frac{E_F-E}{T}  \!\!\left(\frac{\partial A_{\alpha\beta}}{\partial \theta_\gamma }  \!+\!\frac{\partial A_{\alpha\gamma}} {\partial \theta_\beta} 
\!+\!\delta_{\beta\gamma} \Xi_{\alpha\beta}\!\!\right)\!\partial_E f,\\
M_{\alpha\beta\gamma}&=\!\frac{e^2}{h} \!\int\! dE \left(\!\frac{E_F-E}{eT} \frac{\partial A_{\alpha\gamma}}{\partial V_\beta }  \!-\! \frac{\partial A_{\alpha\beta}}{\partial \theta_\gamma} \!-\!\delta_{\beta\gamma} \Xi_{\alpha\beta}\!\right)\partial_E f,
\end{align}
\end{subequations}
where $\Xi_{\alpha\beta}=[(E-E_F)/T] \partial_E A_{\alpha\beta}$.
Notably, the nonlinear responses depend on how the scattering matrix
changes, through the potential $U$, in response to a shift in voltage or temperature.
Since we are concerned with small changes away from equilibrium,
an expansion of $U$ up to first order suffices:
\begin{equation}\label{eq_u}
U=U_\text{eq} + \sum_\alpha u_\alpha V_\alpha +\sum_\alpha z_\alpha \theta_\alpha\,,
\end{equation}
where $u_\alpha=(\partial U/\partial V_\alpha)_\text{eq}$ and $z_\alpha=(\partial U/\partial \theta_\alpha)_\text{eq}$
are characteristic potentials that describe the internal change of the system
to a shift of voltage and temperature, respectively, applied to terminal $\alpha$.
In the sequel, we derive the self-consistent procedure to determine the electrostatic potential
in the presence of electrical and thermal forces. 

The net charge response of the system away from its 
equilibrium state can be decomposed
into two terms, namely, the {\em bare} charge injected 
from lead $\alpha$ and the screening charge that builds up in the 
conductor due to interaction with the injected charges: 
$q=q_\text{bare}+q_\text{scr}$. The contribution to 
$q_\text{bare}$ due to a voltage imbalance in lead $\alpha$ is 
given by the particle injectivity $\nu^{p}_{\alpha}(E)$. This is a partial density of states 
associated with scattering states that describe those carriers originated from lead $\alpha$ \cite{but93}. 
In addition, a shift of temperature in lead $\alpha$ also
induces a change in $q_\text{bare}$. 
In contrast to the voltage case, however, where every carrier with an energy $E$
contributes positively to $q_\text{bare}$,
in the thermally bias case the contribution of a temperature 
shift in lead $\alpha$ gives rise to
a heat addition or removal depending on whether the carrier 
energy $E$ is larger or smaller than $E_F$ \cite{hum02}.
This crucial fact must be reflected in the 
{\em entropic} injectivity denoted by $\nu^{e}_{\alpha}$:
\begin{eqnarray}\label{eq_dpde}
\nu^{p}_\alpha(E)&=&\frac{1}{2\pi i}\sum_{\beta}
\Tr\left[ s^\dagger_{\beta\alpha}\frac{d s_{\beta\alpha}}{dE} \right]\,,\\ \,
\nu^{e}_\alpha(E)&=&\frac{1}{2\pi i}\sum_{\beta}\Tr\left[\frac{E-E_F}{T}
 s^\dagger_{\beta\alpha}\frac{d s_{\beta\alpha}}{dE} \right]\,.\label{eq_dpde2}
\end{eqnarray}
To be concise, we have assumed that the potential is homogeneous (i.e., position-independent)
within the sample (the extension to inhomogeneous fields is straightforward \cite{chr96})
and that the WKB approximation applies in order to make the replacement
$\delta/\delta U\to -e\partial/\partial E$.
We note that the factor $(E-E_F)/T$ represents the entropy transfer
associated to adding a single carrier \cite{emi06}.
Then the accumulation or depletion bare charge imbalance due to voltage or to temperature shifts becomes
$q_\text{bare}=e\sum_\alpha (D^{p}_\alpha eV_\alpha+D^{e}_\alpha \theta_\alpha )$ where $D_\alpha^{p}=-\int dE \nu^{p}_\alpha (E)\partial_E f$, and $D_\alpha^{e}=-\int dE \nu^{e}_\alpha (E)\partial_E f$.
Next, we obtain the screening charge from the response of the internal potential, $\Delta U=U-U_\text{eq}$,
to changes in the leads' chemical potential and temperature. Within the random phase approximation,
one has $q_\text{scr}=e^2\Pi \Delta U$.
$\Pi$ is the Lindhard function which in the static case (frequency-dependent effects are not considered here)
and in the long wavelength limit reads $\Pi=-\sum_\alpha D^{p}_\alpha = - D$ at $T=0$
[$D=D(E_F)$ is the sample density of states] \cite{smi83}. These approximations are excellent for our purpose
since (i) if $T\neq 0$ one can simply replace the previous expression with $\Pi=\int dE D(E) \partial_E f$
and (ii) the long wavelength limit amounts to carrier energies well below the tunnel barrier heights that
couple the conductor to the external reservoirs. But this is precisely the range of validity
of the WKB approximation used to express $D^p$ and $D^e$ in terms
of energy derivatives only.

Our set of equations is closed when we relate the out-of-equilibrium net charge with $\Delta U$ employing the Poisson equation, $\nabla^2 \Delta U=-4\pi q$.
We use Eq.\ \eqref{eq_u} and the fact that $V_\alpha$ and $\theta_\alpha$ shifts are independent. We then identify a pair
of separated equations,
$\nabla^2 u_\alpha + 4\pi e^2 \Pi u_\alpha =- 4\pi e^2 D^{p}_\alpha$ and
$\nabla^2 z_\alpha + 4\pi e^2 \Pi z_\alpha =- 4\pi e D^{e}_\alpha$, which
become nonlocal in the case of inhomogeneous fields.

The voltage and temperature derivatives, $\partial_{\theta_\gamma} A_{\alpha\beta}$ and $\partial_{V_\gamma} A_{\alpha\beta}$,
can be determined once the characteristic potentials are known since
$\partial_{\theta_\gamma} A_{\alpha\beta}=z_\gamma\delta A_{\alpha\beta}/\delta U\to -e z_\gamma\partial_E A_{\alpha\beta}$
and $\partial_{V_\gamma} A_{\alpha\beta}=u_\gamma\delta A_{\alpha\beta}/\delta U\to -e u_\gamma\partial_E A_{\alpha\beta}$.
Thus, Eq. (\ref{nonlinearcoefficients}) becomes
\begin{subequations}\label{eqglm2}
\begin{align}
G_{\alpha\beta\gamma}&=\frac{e^3}{h}\!\!\int \!\!dE\!
\left[ \partial_E A_{\alpha\gamma}u_\beta+\partial_E A_{\alpha\beta}\left(u_\gamma-\delta_{\beta\gamma}\right)\right]\!
\partial_E f\,,\\
L_{\alpha\beta\gamma}&=\frac{e^2}{h}\!\!\int\!\! dE \!
\left[\Xi_{\alpha\gamma} z_\beta+
\Xi_{\alpha\beta} \left( z_\gamma -\frac{E-E_F}{eT}\delta_{\beta\gamma}\right)\right]\!\!\partial_E f\,,\\
M_{\alpha\beta\gamma}&=\frac{e^2}{h}\!\!\int\! \!dE \!
\left[e\partial_E A_{\alpha\beta}z_\gamma+\Xi_{\alpha\gamma}u_\beta-\Xi_{\alpha\beta}\delta_{\beta\gamma}\right]\!\partial_E f\,,
\end{align}
\end{subequations}
This is our central result. Importantly, Eq.\ \eqref{eqglm2} is not only of formal interest but offers clearly practical advantages.
\begin{figure}
  \centering
 \includegraphics[width=0.45\textwidth, clip]{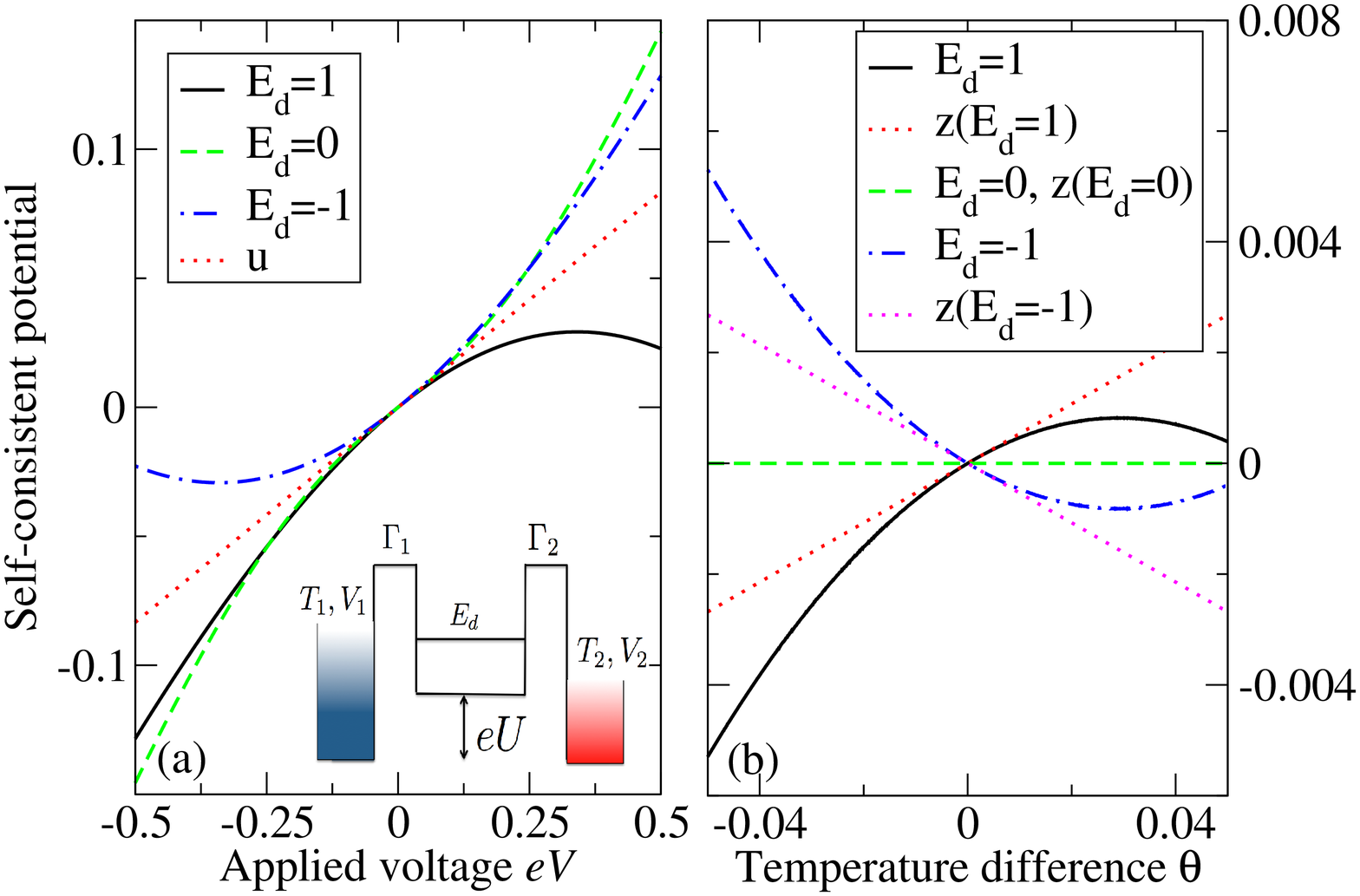}
\caption{(Color online)  Self-consistent screening potential $U$
for a quantum dot system with $V_1=V/2$, 
$V_2=-V/2$ ($E_F=0$), $T_1=T+\theta/2$, and 
$T_2=T-\theta/2$ (see inset). We take $k_B=e=h=1$ and $E_F=0$.
(a) $U$ at $\theta=0$ for $T=0.01$, $\Gamma_1=2\Gamma_2=0.2$
and various dot level positions $E_d$.
The red dotted line corresponds to the leading order
approximation $U=uV=\eta V/2$ independently of $E_d$. 
(b)  $U$ at $V=0$ for $T=0.05$, $\Gamma_2=2\Gamma_1=0.2$ and $E_d=\pm 1,0$.
Colored dotted lines correspond to $U=z\theta$ with the characteristic
potential calculated from Eq.\ \eqref{equz}.}
  \label{fig:1}
\end{figure}

\paragraph{Quantum dot.} 
As an illustrative application of the formalism exposed 
above, we now investigate the nonlinear thermoelectric transport
properties of a quantum dot when Coulomb interactions are treated within a mean-field approximation.
Preliminary observations suggest interesting nonlinear thermoelectric effects in quantum dots \cite{sve12}.
We consider a single level with energy $E_d$ coupled to two reservoirs ($1$ and $2$)
via tunnel barriers (see inset of Fig.\ \ref{fig:1}).
Thus, the level acquires a broadening given by $\Gamma=\Gamma_1+\Gamma_2$.
The corresponding Breit-Wigner lineshape depends, quite generally, on the internal potential $U$, 
which is self-consistently calculated through the Poisson equation.  
The dot charge is then
\begin{equation}\label{q_dot}
q_d=\frac{e}{\pi }\int dE\frac{\Gamma_1 f_1(E)+
\Gamma_2 f_2(E)}{(E-E_d-e U)^2 +\Gamma^2 } \,.
\end{equation}
We expand Eq. (\ref{q_dot}) to leading order in $V_\alpha$, $\theta_\alpha$ and $U$.
We find $\delta q_d =  e^2 D^{p}_1 V_1+ e^2 D^{p}_2 V_2 + e 
D^{e}_{1}  \theta_1 + e D^{e}_{2} \theta_2 - e^2 D U$,
where $\delta q_d=q_d-q_d^e$ denote the charge excess due to voltage and temperature shifts and
$q_d^e$  is the equilibrium charge given by Eq. (\ref{q_dot}) with $f_1=f_2=f$.
$D^{p}_{\alpha}=-\frac{\Gamma_\alpha}{\pi}\int dE \frac{1}{(E-E_d)^2 +\Gamma^2}  \partial_E f$
and $D^{e}_{\alpha}=-\frac{\Gamma_\alpha}{\pi}\int dE \frac{E-E_F}{T} \frac{1}{(E-E_d)^2 +\Gamma^2}  \partial_E f$
are the integrated particle and entropic injectivities
of Eqs.\ \eqref{eq_dpde} and \eqref{eq_dpde2} when the Breit-Wigner representation is used.

In a discrete form, the Poisson equation
 is written in terms of a geometrical capacitance 
 $C$ which connects electrically the dot to an external gate terminal.
Accordingly, the charge excess of the dot obeys
$\delta q_d=C (U-V_g)$ where $V_g$ is the gate potential. Then,
\begin{equation}\label{internal_U}
U=\frac{e^2 D^{p}_1 V_1+ e^2 D^{p}_2 V_2 + 
e D^{e}_{1}  \theta_1 + e D^{e}_{2} \theta_2 +C V_g}{C+e^2 D}\,,
\end{equation}
from which the characteristic potentials follow,
\begin{equation}\label{equz}
u_{1(2)}=\frac{e^2 D^p_{1(2)}}{C+e^2 D},\,\,u_g=\frac{C}{C+e^2 D},\,\,
z_{1(2)}=\frac{e D^e_{1(2)}}{C+e^2 D}\,.
\end{equation}

\paragraph{Rectification effects.} We consider the charge neutral limit ($C=0$)
since it applies to the experimentally relevant case
of strong interactions. Moreover, if the dot is 
symmetrically biased
($V_1=V/2$, $V_2=-V/2$, $T_1=T+\theta/2$, and $T_2=T-\theta/2$) then
$u=\partial U/\partial V=\eta/2$ and $z=\partial U/\partial \theta=(D_1^e-D_2^e)/[2e (D_1^p+D_2^p)]$
to leading order in $V$ and $\theta$
with $\eta=(\Gamma_1-\Gamma_2)/\Gamma$ the tunneling asymmetry \cite{san04}.
In Fig.\ \ref{fig:1} we show the exact dot potential obtained from a numerical calculation
of Eq.\ \eqref{q_dot} compared to its approximate value [Eq.\ \eqref{internal_U}].
We distinguish between the isothermal case [$\theta=0$, Fig. \ref{fig:1}(a)]
and the isoelectric case [$V=0$, Fig. \ref{fig:1}(b)].
In the former, the self-consistent potential is plotted for three values of the dot level $E_{d}=\pm 1,0$.
The curves for the exact $U$
agree with approximation $U=u V$ at low $V$, as expected.
In the strongly nonlinear regime and for $E_d=\pm 1$, higher order terms ($V^2$ or
higher) make $U$ depart from its linearity. We recall
that linear responses depend on $U_{\rm eq}$ only and they are insensitive
to the variation of $U$ with $V$. Only the nonlinear current allows us to explore this regime.
Interestingly, at resonance ($E_d=0$) the contributions to $U$ from even powers in $V$ are absent.
In the isoelectric case [Fig. \ref{fig:1}(b)], we present
$U$ in response to a thermal shift for $E_d=\pm 1,0$. Particularly interesting is the
particle-hole symmetry case $E_d=0$ for which $U$ vanishes to all $\theta$ powers. 
We also compare the full calculation with the leading-order approximation $U=z\theta$.
Notice that contrary to the isoelectric case $z$ depends on $E_d$. 
The agreement is quite reasonable at low temperature shifts.
 \begin{figure}
  \centering
 \includegraphics[width=0.45\textwidth,clip]{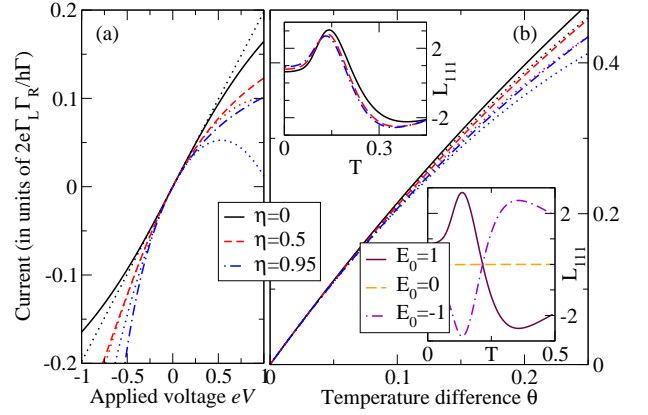}
  \caption{(Color online)  Electrical current
for a quantum dot system with $V_1=V/2$, 
$V_2=-V/2$, $T_1=T+\theta$ and 
$T_2=T$ for $\Gamma=0.2$ and three different values
of the tunneling asymmetry $\eta$. (a) $I$--$V$ characteristics for $\theta=0$ and $T=0.01$ along with 
the leading-order nonlinearity $I\simeq G_{11}V+G_{111} V^2$.
The latter correspond to the colored dotted lines. (b) $I$--$\theta$ characteristics for $V=0$ and $T=0.5$.
Colored  dotted lines correspond to $I\simeq L_{11} \theta+L_{111}\theta^2$. Upper inset: 
$L_{111}$ versus $T$ as a function of $\eta$ for $E_d=1$. Lower inset: $L_{111}$ for $E_d=\pm 1,0$ and $\eta=0$.}
  \label{fig:2}
\end{figure}

The evolution of the current for an electrically and 
thermally driven quantum dot is shown in Fig. \ref{fig:2}(a) 
and Fig. \ref{fig:2}(b), respectively, for fixed $E_d=1$. For $\theta=0$
the current first follows Ohm's law at low $V$ and then, at higher voltages,
acquires a $V^2$ dependence leading to rectification effects.  
The $I$--$V$ curves can be approximated up to $V^2$ with  $I= G_{11} 
V+G_{111}V^2+\mathcal{O}(V^3)$ where the leading-order nonlinearity
in the Sommerfeld approximation, $G_{111}= \frac{e^3}{h} \partial_E A_{11}|_{E=E_F}(1- 2u_1)$,
depends on the internal potential response. The $I$--$V$ curves in Fig. \ref{fig:2}(a) 
correspond to three values of $\eta$
and show good agreement with the second order expansion except 
for very high voltages.
In Fig. \ref{fig:2}(b) we show $I$ driven by a temperature shift for $V=0$.  
We  compare the full $I$--$\theta$ characteristics for different $\eta$ values
  with the second-order expansion,
 $I= L_{11} \theta+L_{111} \theta^2+\mathcal{O}(\theta^3)$,
 where the thermal rectification term is
 \begin{equation}\label{eq_L111}
 L_{111}=\frac{e\pi^2 k_B^2}{3h} (\partial_E A_{11}|_{E=E_F}- 2 e z_1 T\partial_E^2 A_{11}|_{E=E_F}) \,,
 \end{equation}
 to leading order in the Sommerfeld approximation.
First, $I$ grows linearly with $\theta$ and then higher orders in $\theta$
 become relevant above a threshold where $L_{111}$
 is large enough.  We plot $L_{111}$ in Fig. \ref{fig:2}(b) (upper inset)
 and find a nonmonotonic behavior with the background temperature $T$. 
We also show Fig. \ref{fig:2}(b) (lower inset)
the dependence of $L_{111}$ for various level positions
and $\eta=0$. 
Interestingly, in the particle-hole  symmetry point $L_{111}$ vanishes
 identically (like $L_{11}$) whereas for $E_d=\pm 1$, $L_{111}$
 presents an opposite behavior as a function of $T$. It also follows from Eq. \eqref{eq_L111}
 that for $T=0$ $L_{111}$ is generally nonzero unlike $L_{11}$.
\begin{figure}
  \centering
 \includegraphics[width=0.45\textwidth, clip]{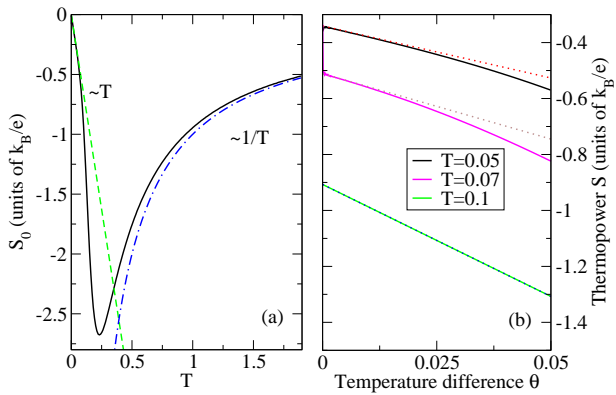}
  \caption{(Color online) (a) Linear-response thermopower $S_0$ for a symmetrically voltage biased dot and
  one heated contact ($\theta_1=\theta$ and $\theta_2=0$) at $E_d=1$. Low and high temperature limits of
 $S_0$ are explicitly shown. (b) Thermopower $S$ beyond linear response
 for three different background temperature values. We show with colored dotted lines the leading-order
 expansion $S\simeq S_0+S_1 \theta$ calculated from the sensitivity given by Eq.\ \eqref{eqs1}.}
  \label{fig:3}
\end{figure}

\paragraph{Thermopower.} The thermopower $S$ yields the voltage generated across
the sample in response to an applied thermal bias at vanishing current condition.
In the linear transport regime and for a two-terminal conductor, the Seebeck
coefficient is $S_0=V/\theta|_{I=0}=-L_{11}/G_{11}$. This expression is correct
in the limit $\theta\to 0$. At low temperatures,
it can be approximated to the Mott formula $S_{0}\simeq - (\pi^2 k_B^2 T/3e) \partial_E \ln A_{11}|_{E=E_F}\propto T$
whereas for high $T$ we find $S_{0}\simeq (E_F-E_d)/eT\propto T^{-1}$ in the limit $\Gamma\ll k_B T$. In Fig.\ \ref{fig:3}(a)
we numerically calculate $S_0$ for an electrically biased quantum dot
($V_1=-V_2=-V/2$) when only one reservoir is heated ($\theta_1=\theta$ and $\theta_2=0$). 
Our numerical simulations reproduce the analytical $T$-dependence both at low temperature
(Mott relation) and at high temperature (infinitely narrow resonance).
More interesing are the $\theta$-corrections to $S$ when $\theta$ is not small.
Then, we can expand $S= S_0+S_1 \theta+\mathcal{O}(\theta^2)$
where the $S_1$ is the thermopower {\em sensitivity} which measures
the deviations of $S$ from a constant value. Importantly, a measurement
of the differential thermopower $d S/d\theta$ gives precisely $S_1$
to leading order in $\theta$. Specializing Eq.\ \eqref{I_exp} to the two-terminal
case and setting $I=0$ we find
\begin{equation}\label{eqs1}
S_{1}=-\frac{1}{G_{11}^3}[G_{111} L_{11}^2+ L_{111}G_{11}^2+G_{11}L_{11}(M_{121}-M_{111})]\,,
\end{equation}
valid when a single lead is heated. 
Inserting Eq.\ \eqref{nonlinearcoefficients} in Eq.\ \eqref{eqs1},  we compare the sensitivity
with an exact calculation of $S$ for a quantum dot as above.
We observe in Fig.\ \ref{fig:3}(a) that excellent agreement is found for
low $\theta$ and that departures depend on the particular value of $T$.
It is also noteworthy that in the low $T$ limit the second term in brackets of Eq.\ \eqref{eqs1}
dominates since $L_{11}\propto T^2$ and $L_{11}(M_{121}-M_{111})\propto T^2$ within a Sommerfeld
expansion. Then, according to Eq.\ \eqref{eq_L111} a low-temperature measurement of
the thermopower sensitivity would provide information on the renormalization of the dot leve due to
a temperature gradient.

\paragraph{Conclusions.}
We have presented a general nonlinear scattering theory 
for mesoscopic conductors that are driven by electrical and 
thermal gradients. In the weakly nonlinear regime,
screening effects arise in response to charge pile-up
due to voltage or temperature differences. Importantly,
the transmission probability becomes a function of the
thermal gradient. We have found
that the screening response
can be described in terms of particle {\em and}
entropic injectivities. We have illustrated our theory
with an application to a two-terminal quantum dot setup,
evaluating the current--voltage and current--temperature
characteristics. Importantly, we have discussed
thermopower sensitivity in the nonlinear regime of transport.
Our results are relevant in view of recent
advances in thermoelectrics at the nanoscale.

\paragraph{Acknowledgments.}
We thank M. B\"uttiker and H. Linke for useful suggestions.
Work supported by MINECO Grant No. FIS2011-23526.

\end{document}